\documentclass[aps,twocolumn,amsmath,amssymb]{revtex4}
\usepackage{graphicx}
\usepackage{amsmath}
\usepackage{amssymb}

\newcommand{\kpe}{\mathbf{k}\!\cdot\!\mathbf{p}}

\newcommand{\tn}[1]{\textnormal{#1}}
\begin{document}
\title{Polarized Emission Lines from Single InGaN/GaN Quantum Dots: Role of the Valence-band Structure of Wurtzite Group-III Nitrides}
\author{M.~Winkelnkemper\footnote{Also at: Fritz-Haber-Institut der Max-Planck-Gesellschaft, Faradayweg 4-6, D-14195 Berlin, Germany}} 
 \email{Momme@sol.Physik.TU-Berlin.de}
\affiliation{Institut f\"ur Festk\"orperphysik, Technische
Universit\"at Berlin, Hardenbergstra{\ss}e 36, D-10623 Berlin, Germany}
\author{S.~Seguin}
\affiliation{Institut f\"ur Festk\"orperphysik, Technische Universit\"at Berlin, Hardenbergstra{\ss}e 36, D-10623 Berlin, Germany}
\author{S.~Rodt}
\affiliation{Institut f\"ur Festk\"orperphysik, Technische
Universit\"at Berlin, Hardenbergstra{\ss}e 36, D-10623 Berlin, Germany}
\author{A.~Schliwa\footnote{Current address: Weierstrass Institut f\"ur Angewandte Analysis und Stochastik, D-10117 Berlin, Germany}
}
\affiliation{Institut f\"ur Festk\"orperphysik, Technische
Universit\"at Berlin, Hardenbergstra{\ss}e 36, D-10623 Berlin, Germany}
\author{L.~Rei{\ss}mann\footnote{Current address: Institut f\"ur Experimentelle Physik, Otto-von-Guericke-Universit\"at Magdeburg, D-39016 Magdeburg, Germany}
}
\affiliation{Institut f\"ur Festk\"orperphysik, Technische
Universit\"at Berlin, Hardenbergstra{\ss}e 36, D-10623 Berlin, Germany}
\author{A.~Strittmatter}
\affiliation{Institut f\"ur Festk\"orperphysik, Technische
Universit\"at Berlin, Hardenbergstra{\ss}e 36, D-10623 Berlin, Germany}
\author{A.~Hoffmann}
\affiliation{Institut f\"ur Festk\"orperphysik, Technische
Universit\"at Berlin, Hardenbergstra{\ss}e 36, D-10623 Berlin, Germany}
\author{D.~Bimberg}

\affiliation{Institut f\"ur Festk\"orperphysik, Technische
Universit\"at Berlin, Hardenbergstra{\ss}e 36, D-10623 Berlin, Germany}

\date{\today}
\begin{abstract}
We present a study of the polarization properties of emission lines from single InGaN/GaN quantum dots (QDs).  The QDs, formed by spinodal decomposition within ultra-thin InGaN quantum wells, are investigated using single-QD cathodoluminescence (CL). The emission lines exhibit a systematic linear polarization in the orthogonal crystal directions [$11\overline{2}0$] and [$\overline{1}100$]---a symmetry that is non-native to hexagonal crystals.

Eight-band $\kpe$ calculations reveal a mechanism that can explain the observed polarizations: The character of the hole(s) in an excitonic complex determines the polarization direction of the respective emission if the QD is slightly elongated.  Transitions involving $A$-band holes are polarized parallel to the elongation; transitions involving $B$-type holes are polarized in the orthogonal direction.  The energetic separation of both hole states is smaller than 10\,meV.  The mechanism leading to the linear polarizations is not restricted to InGaN QDs, but should occur in other wurtzite-nitride QDs and in materials with similar valence band structure.

\noindent
\textsl{(Conf.\ proc.\ of the MSS-13 in Genova 2007, to be publish in Physica E. \copyright Elsevier)}  
\end{abstract}
\maketitle
Recently, single-photon emitters (SPEs) operating at 200\,K---a temperature that can easily be achieved by Peltier cooling---have been realized employing nitride quantum dots (QDs) \cite{kako2006}.  In addition, the emission of linearly polarized photons is essential for successful implementation of quantum-cryptographic protocols \cite{BB84}.

We have shown recently that emission lines from single InGaN/GaN QDs are systematically linearly polarized in the orthogonal crystal directions [$11\overline{2}0$] and [$\overline{1}100$]---a symmetry that is non-native to hexagonal crystals \cite{winkelnkemper2007}.   The polarization of the emission lines has been attributed to the character of the hole state [formed by the $A$ or $B$ valence band (VB)] involved in the recombination process and a slight elongation of the QDs.

In this Article, we address this polarization mechanism.  Its origin in the VB structure of group-III nitrides and its dependence on the degree on elongation will be analyzed.

The sample structure and experimental details have been described in-depth elsewhere \cite{winkelnkemper2007,seguin2004}. Briefly, the sample was grown in the wurtzite phase by metal-organic chemical vapor deposition on Si(111) substrate. The InGaN layer was grown at 800\,$^{\circ}$C with a nominal thickness of 2\,nm using trimethylgallium, trimethylindium, and ammonia as precursors. The QDs form within the InGaN-layer due to spinodal decomposition.  The sample has been investigated with single-QD cathodoluminescence employing Pt shadow masks to increase the spatial resolution.  Further discrimination of the single-QD lines was possible using their temporal jitter: All lines originating from the same QD show the same characteristic jitter pattern.  Line groups of up to five narrow lines with the same jitter pattern could be identified \cite{winkelnkemper2007}.  An example of such a single-QD spectrum is shown in Fig.~1. All emission lines show a pronounced linear polarization. The polarization directions scatter around [$\overline{1}100$] and [$11\overline{2}0$]. Both directions were found in each line group \cite{winkelnkemper2007}.  

The emission patterns differ significantly from well-known polarized emission spectra of II/VI and III/V QDs with cubic crystal structure, where the polarization is caused by, e.g., the fine-structure splitting of the exciton (see, e.g., Refs.~\onlinecite{bayer2002,kulakovskii1999,seguin2005}).  
\begin{figure}
\centering
\includegraphics[width=0.9\columnwidth]{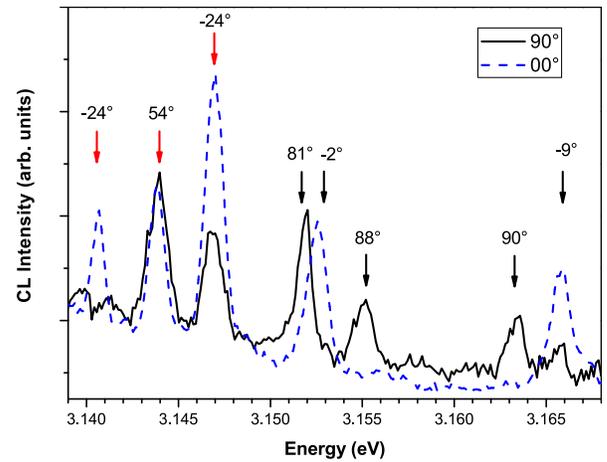}
\caption{\label{fig:cl}Sample of polarized spectra of two QDs (red and black arrows). 85° (-5°) degrees corresponds to the [$\bar{1}$100] ([11$\bar{2}$0]) crystal direction. The numbers give the polarization direction of each line as obtained from the analyses of spectra with different polarization angles. Lines from one QD are polarized perpendicular to
each other.}
\end{figure}

For the calculations we use an eight-band $\kpe$ model, implemented for arbitrarily shaped QDs on a three-dimensional finite differences grid \cite{winkelnkemper2006}. The model includes strain effects, piezo- and pyroelectricity, conduction-band(CB)/VB coupling, and crystal-field and spin-orbit splitting. Recently developed optimized $\kpe$ parameters \cite{Rinke2006} have been included as described in Ref.~\onlinecite{winkelnkemper2007}.
In accord with recent publications \cite{winkelnkemper2007,winkelnkemper2006}, the QDs have been modeled as ellipsoids with linearly graded indium concentration. The maximum indium concentration at the QDs' centers $x_\tn{c}$ is $50$\,\%, the minimum concentration at the QDs' borders $x_\tn{e}$ is $5$\,\%. The QDs are embedded in an $\tn{In}_{0.05}\tn{Ga}_{0.95}\tn{N}$ quantum well with a height of $2$\,nm. The QDs have a height of $d_z=2$\,nm and lateral extensions of $d_{x/y}=4.6-5.8$\,nm. Henceforth $y$ will denote the direction of the QDs long axis.
\begin{figure}
\centering
\includegraphics[width=\columnwidth]{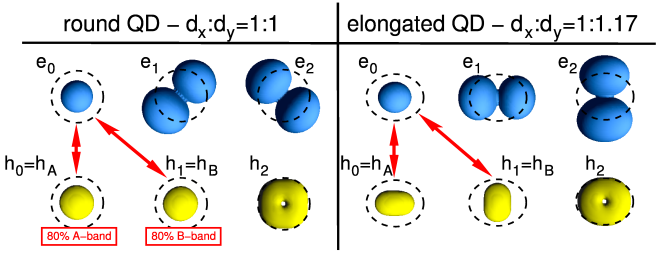}
\caption{\label{fig:states}Single-particle states in a round (left) and an elongated (right) QD.of transitions between the electron and hole states.}
\end{figure}

The emission lines can be attributed to transitions involving either the confined hole ground state ($h_0\equiv h_A$), which is predominantly formed by the $A$ VB, or the first excited hole state ($h_1\equiv h_B$), which is predominantly formed by the $B$ VB (see Tab.~\ref{tab:proj}).  If a QD is slightly elongated, transitions involving $h_A$ are polarized parallel to the elongation and transitions involving $h_B$ perpendicular to it \cite{winkelnkemper2007}.

\begin{table}
\centering
\caption{\label{tab:proj} Projections of the hole states $h_A$ and $h_B$ on the bulk $A$ and $B$ VBs and on the $\kpe$ basis states $P_x$ and $P_y$.}

\begin{tabular}{l|cccc|cccc}
\hline
\hline
& \multicolumn{4}{c}{$h_0\equiv h_A$} & \multicolumn{4}{c}{$h_1\equiv h_B$}\\
$d_\tn{y}/d_\tn{x}$ & $A$ & $B$ & $P_x$ & $P_y$& $A$ & $B$ & $P_x$ & $P_y$\\
\hline
$1$ & 0.81 & 0.13 & 0.47 & 0.47 & 0.16 & 0.78 & 0.47 & 0.47 \\
$1.08$ & 0.76 & 0.18 & 0.26 & 0.68 & 0.21 & 0.73 & 0.67 & 0.26 \\
$1.17$ & 0.69 & 0.25 & 0.16 & 0.78 & 0.28 & 0.66 & 0.77 & 0.17 \\
$1.25$ & 0.66 & 0.29 & 0.12 & 0.83 & 0.32 & 0.62 & 0.81 & 0.13 \\
\hline
\hline
\end{tabular}
\end{table}
The single-particle electron and hole orbitals are depicted in Fig.~2 (upper part) for a circular QD (left) and a slightly elongated QD (right).  The electron ground-state ($e_0$) envelope functions have $s$-like symmetry; the ones of the first two excited electron states ($e_{1/2}$) are $p$-like.  The first two hole states, $h_A$ and $h_B$, in contrast, both have $s$-like envelope functions.  Both states have sizable oscillator strengths with the electron ground state, but behave differently if the QD is elongated: $h_A$ aligns parallel to the QD's long axis, $h_B$ perpendicular to it.  An analysis of the projections of both hole states on the $\kpe$ basis states $P_x$ and $P_y$ \cite{chuang1996} yields that the $P_y$ ($P_x$) projection of $h_A$ ($h_B$) increases with increasing QD elongation, while the $P_x$ ($P_y$) projection decreases (see Tab.~\ref{tab:proj}). Consequently, the optical transition between $h_A$ ($h_B$) and $e_0$ is linearly polarized parallel (perpendicular) to the QD's long axis.

\begin{figure}
\centering
\includegraphics[width=\columnwidth,clip]{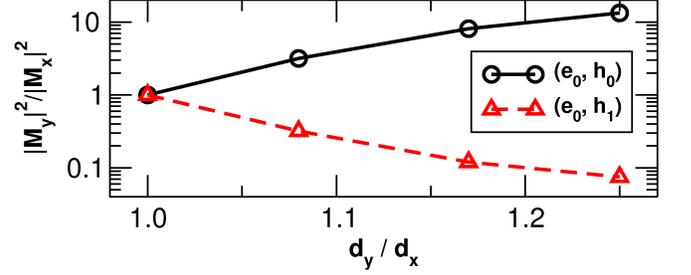}
\caption{\label{fig:poldegree} Degree of polarization of the transitions between $e_0$ and $h_A$ or $h_B$, respectively, for different degrees of QD elongation.}
\end{figure}

Figure~3 shows the ratio $|M_y|^2/|M_x|^2$ of the optical matrix elements ($M_i$) in $y$ and $x$ direction for the $A$- and the $B$-band transition.  Even for very slight elongations, a pronounced polarization is found: For the model QD with the slightest anisotropy ($d_y/d_x=1.08$) the polarization degree is already $\approx 3$:$1$. Upon further elongation of the QDs it increases drastically.

Hence, the polarization of optical transitions in wurtzite InGaN QDs can be controlled easily on the basis of the properties of the single-particle hole states, which are sensitive to the QD shape. The mechanism should also apply to other wurtzite-nitride QDs, like GaN/AlN QDs, due to its origin in the VB structure of wurtzite group-III nitrides.  

In summary, we have presented a study of the polarization properties of emission lines from single InGaN/GaN QDs. Experimentally, a systematic polarization of the emission lines from single InGaN/GaN QDs has been observed. The polarization has been explained theoretically with transitions involving either $A$- or $B$-band hole states and a slight elongation of the QDs. This mechanism can be exploited for generation of polarization-based photonic qubits using nitride QDs.  

This work was in part funded by Sfb 296 of DFG and SANDiE Network of Excellence (No.\ NMP4-CT-2004-500101) of the European Commission. Parts of the calculations were performed on the IBM pSeries 690 supercomputer at HLRN within Project No.\ bep00014.

\end{document}